# Topological phase singularities in atomically thin high-refractive-index materials

*Georgy Ermolaev[†],[1] Kirill Voronin[†],[1,2] Denis G. Baranov,[1] Vasyl Kravets,[3] Gleb Tselikov,[1] Yury Stebunov,[4] Dmitry Yakubovsky,[1] Sergey Novikov,[1] Andrey Vyshnevyy,[1] Arslan Mazitov,[1,5] Ivan Kruglov,[1,5] Sergey Zhukov,[1] Roman Romanov,[6] Andrey M. Markeev,[1] Aleksey Arsenin,[1] Kostya S. Novoselov,[4,7,8] Alexander N. Grigorenko,[3] and Valentyn Volkov[1,\*]*

[1]Center for Photonics and 2D Materials, Moscow Institute of Physics and Technology, Dolgoprudny 141700, Russia.

[2]Skolkovo Institute of Science and Technology, Moscow 121205, Russia.

[3]Department of Physics and Astronomy, University of Manchester, Manchester M13 9PL, UK.

[4]National Graphene Institute (NGI), University of Manchester, Manchester M13 9PL, UK.

[5]Dukhov Research Institute of Automatics (VNIIA), Moscow 127055, Russia.

[6]National Research Nuclear University MEPhI (Moscow Engineering Physics Institute), Moscow 115409, Russian Federation.

[7]Department of Materials Science and Engineering, National University of Singapore, Singapore 03-09 EA, Singapore.

[8]Chongqing 2D Materials Institute, Chongqing 400714, China.

[\*]e-mail: volkov.vs@mipt.ru

**ABSTRACT**

Atomically thin transition metal dichalcogenides (TMDCs) present a promising platform for numerous photonic applications due to excitonic spectral features, possibility to tune their constants by external gating, doping, or light, and mechanical stability. Utilization of such materials for sensing or optical modulation purposes would require a clever optical design, as by itself the 2D materials can offer only a small optical phase delay – consequence of the atomic thickness. To address this issue, we combine films of 2D semiconductors which exhibit excitonic lines with the Fabry-Perot resonators of the standard commercial $SiO_2$/Si substrate, in order to realize topological phase singularities in reflection. Around these singularities, reflection spectra demonstrate rapid phase changes while the structure behaves as a perfect absorber. Furthermore, we demonstrate that such topological phase singularities are ubiquitous for the entire class of atomically thin TMDCs and other high-refractive-index materials, making it a powerful tool for phase engineering in flat optics. As a practical demonstration, we employ $PdSe_2$ topological phase singularities for a refractive index sensor and demonstrate its superior phase sensitivity compared to typical surface plasmon resonance sensors.

**INTRODUCTION**

Optical waves carry energy and information encoded in their electric field amplitude, phase, and polarization. While light field amplitude is still used the most in various applications, optical phase manipulation could lie in the core of next-generation information technologies.[1–3] Generally, the phase acquired by light upon reflection, scattering, or transmission varies rapidly when the amplitude of the light (reflected, scattered, or transmitted, respectively) goes to zero.[4–7] Reflection zeros can be encountered in the effect of plasmonic blackbody,[8] perfect absorption,[9,10] coherent perfect absorption,[11] Brewster angle[12,13] or more sophisticated examples of zero-reflection modes.[5,14] Such zeros of response function always reveal phase singularities[4,5,15,16] which are accompanied by a non-trivial topological charge $C = \frac{1}{2\pi}\oint_\gamma grad(\varphi)ds$, where $\varphi$ is the response function phase while the integration is performed along a path $\gamma$ enclosing the singular point (zero response point) in two-dimensional parameter space.[5,15,16]

To demonstrate how phase singularities associated with zero-reflection can be topologically protected, let us consider light reflection from a planar structure shown in Figure 1a, where an atomically thin layer of a high-refractive index material (HRIM) is placed on top of $SiO_2$/Si substrate. For a given thickness of HRIM film, an angle of incidence ($\phi$), the photon energy ($E$), and polarization, reflection from the structure can be made exactly zero by calculating appropriate values for the dielectric permittivity of the HRIM film ($Re(\varepsilon)$ and $Im(\varepsilon)$), which can be always found by the nature of Fresnel coefficients for the structure.[17] When we change an angle of incidence and a photon energy in some range, a zero-reflection surface will appear, see the blue surface in Figure 1b. Any point on the zero-reflection surface corresponds to zero-reflection (at some angle of incidence and light wavelength) for the studied structure. Now we plot an actual dependence of the dielectric constants HRIM on the photon energy on the same graph (the so-called material dispersion curve), see the red line in Figure 1b. In the presence of a reasonably large resonance feature, this curve would look like a spiral in the space of ($E$, $Re(\varepsilon)$ and $Im(\varepsilon)$) and hence will inevitably cross the surface of zero-reflection as shown in Figure 1b. For a Lorentz resonance feature,[17] there will be two intersection points between the material dispersion curve and the zero-reflection point resulting in two zeros of reflection from the structure in Figure 1a. It is necessary to stress that these intersection points are protected by the Jordan theorem.[5,6] Indeed, minor variations of the material dispersion curve caused by material imperfections will not change the relative alignment of the curve and the zero-reflection surface and cannot lead to disappearance of the zero-reflection points which lead to the idea of topological darkness.[5,6] Our structure relies on the atomic flatness of the interfaces and on the sharp changes of the refractive indexes, which offer a non-trivial possibility to realize zero-reflection and topological singularities for atomically thin layers in this structure at visible light, which was never achieved before.

Zero-reflection implies perfect absorption of the light that falls onto the discussed structure (as transmission through the silicon substrate should be zero). Zero-reflection entails phase singularity due to the singular nature of light phase at zero light amplitude (where the phase of light is not defined). Figure 1c represents the map of the p-polarized wave reflection phase in space of $E$ and $\phi$, which contains two phase singularities corresponding to the zero-reflection points with topological charges equal to $-1$ and $+1$ (the topological points of Figure 1b). Of immediate interest is a bifurcation behavior of optical phase in the vicinity of topological point (Figure 1d). In other words, phase reveals abrupt $\pm\pi$-jumps near a zero-reflection when plotted as a function of wavelength for a fixed incidence angle close to a phase singularity. It gives an indispensable degree of freedom for efficient phase manipulation.

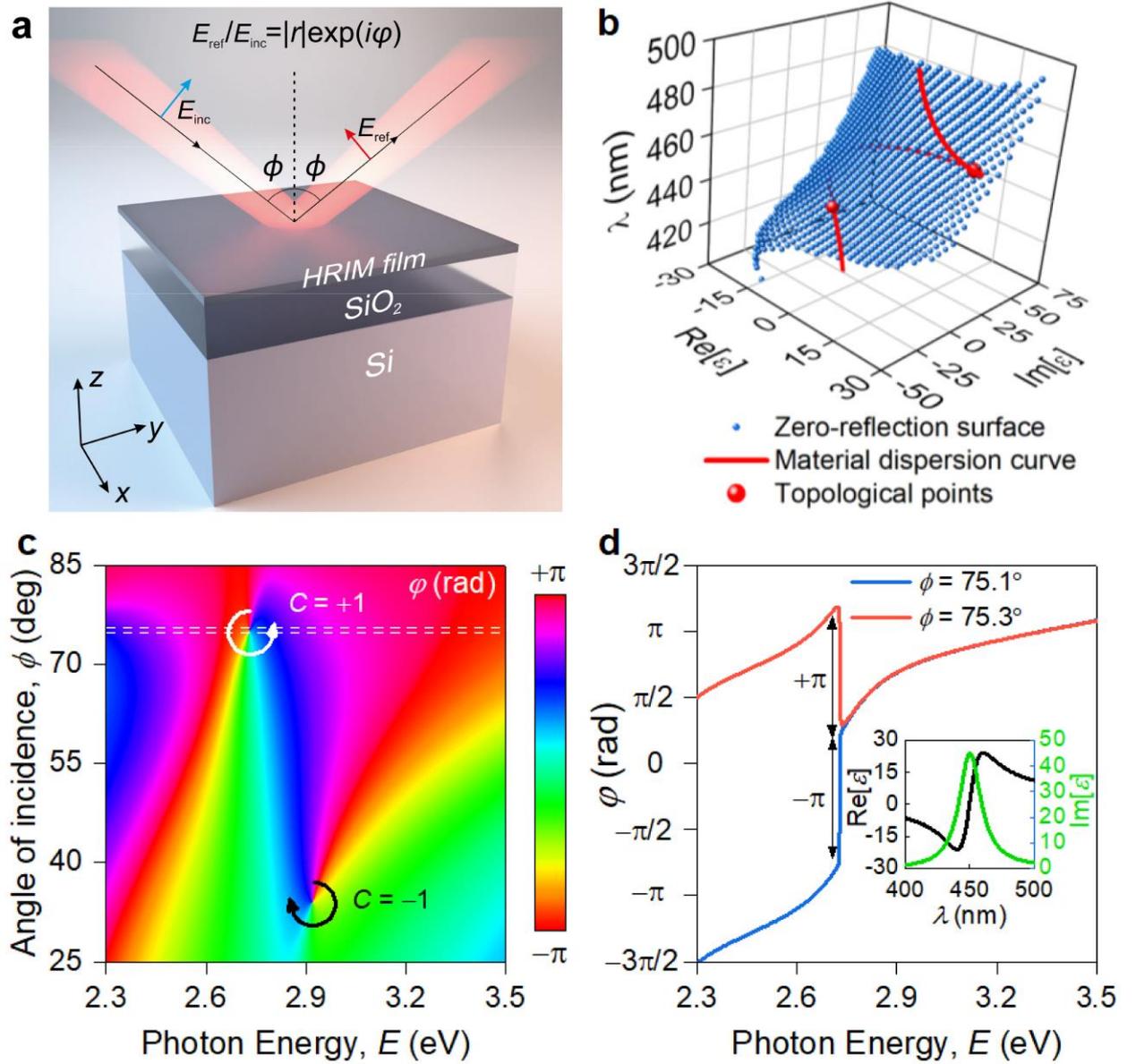

**Figure 1. Topology of the reflection phase near the singular point. a** Schematics of generalized structure for observation of phase singularities, arising from interaction of Fabry-Perot resonator's (280 nm SiO$_2$/Si) modes with ultrathin films of HRIM. **b** Phase singularity point arises when zero-reflection surface of the system HRIM/SiO$_2$/Si intersects with the material dispersion curve of HRIM. **c** In close vicinity of zero-reflection points, phase becomes singular and acquires topological charge $C = -1$ or $+1$, corresponding to –2π or +2π phase round-trip accumulation. **d** Phase has opposite π-gradient for angles slightly above and below singular point (dashed lines in panel (c)) giving rise to topological charge with 2π round-trip around zero-reflection point. The inset is a dielectric permittivity of the model HRIM used for calculation of (b) and (c).

The most exciting consequence of such optical phase control is the realization of "flat optics" paradigm – flexible manipulation of the optical wavefront by an arrangement of subwavelength planar objects to shape the desired phase pattern.[1,18–20] Such "flat optics" paradigm enables miniaturized metalenses[21–23] and meta-holograms,[24–26] and two-dimensional (2D) materials[27] such as graphene,[28] transition metal dichalcogenides,[29] and organic semiconductors[30] provide an excellent platform for implementation of these components. Although these works constitute an important step towards truly flat optics, the efficiency of current devices is limited by the fundamental constraints. Indeed, the typical wavefront manipulation applications require that the phase accumulated by a light wave upon interaction with such a device can be tuned at least within the range of $\pi$. However, in the monolayer limit ($t \sim 0.65$ nm and $n \sim 4$) for visible light ($\lambda \sim 600$ nm), the resulting phase delay, which is approximately determined by the optical thickness of the 2D material layer, is only about $0.01\pi$. Consequently, finding new ways to induce strong optical phase variations in atomically thin structures is vital for flat optics.

Here, we demonstrate a platform for efficient optical phase manipulation presented by atomically thin high-refractive-index materials (HRIMs) that often possess excitonic resonances. We experimentally observe zero-reflection, phase singularities, and rapid phase variation of reflected light in extremely thin layers (down to single monolayer!) of $PdSe_2$, graphene, $MoS_2$, and $WS_2$ films placed on $SiO_2/Si$ substrate. Combined theoretical and experimental analysis indicates that the zero-reflection points are accompanied by a non-trivial topological charge. We derive an analytical condition for such points to occur in layered structures containing optically thin films and predict the occurrence of these points in structures containing a broader family of atomically thin HRIMs and substrates. The observed effect is highly robust and does not require complicated fabrication steps guaranteeing its reproducibility and reliability. In contrast to optical darkness observed for dielectric materials and multilayers which disappears with layer irregularities (e.g., at Brewster angle conditions) the effect is topologically protected. It can be used in numerous applications, including label-free bio- and chemical sensing, photo-detection and photo-harvesting, perfect light absorption in 2D monolayers, quantum communication and security. As a practical application of this platform, we demonstrate a refractive index sensor that can rival modern plasmon resonance-based counterparts.[31] Therefore, our phase engineering approach as a whole offers an advanced tool for current and next-generation 2D flat optics.

**RESULTS**

**Phase singularities in reflection**

To examine topological properties of the light reflectance from thin HRIM films placed on $SiO_2$/Si substrates, we utilized spectroscopic ellipsometry (Methods). The unique advantage of this technique is the simultaneous determination of reflection amplitude and phase in terms of the ellipsometric parameters $\Psi$ and $\Delta$, which are defined through the complex reflection ratio $\rho$:[32]

$$\rho = \tan(\Psi)e^{i\Delta} = \frac{r_\mathrm{p}}{r_\mathrm{s}} \tag{1}$$

where $r_\mathrm{p}$ and $r_\mathrm{s}$ are the amplitude reflection coefficients of p- and s-polarized plane waves. Therefore, ellipsometry provides us with the information not only about the reflected light amplitude, but also about the light phase.

Unexpectedly, we found that light reflection measured from 5.1 nm thick $PdSe_2$ as well as for monolayers of graphene, $MoS_2$, and $WS_2$ on $SiO_2$/Si substrate showed a number of zero-reflection points as explained in Figure 2. Remarkably, the measured amplitude parameter $\Psi$ in Figure 2a is in excellent agreement with the simulated spectrum in Figure 2b calculated using the transfer-matrix method.[33] However, the spectra of $\Psi$ alone do not definitively indicate if an exact zero was attained in reflection at the position of any of $\Psi$ dips. This can be deduced from the behavior of relevant phase which was measured using spectroscopic ellipsometry (phase $\Delta$). The angle-dependent spectrum of the measured ellipsometric phase $\Delta$ in Figure 2c and d clearly indicates that the reflected phase is undefined in the vicinity of certain points in the energy-incidence angle parameter space and possesses non-trivial topological charge $C$. These points are phase singularities, which can occur if and only if the response function (reflection in our case) takes zero magnitude at that point. Figure 2 reveals three such phase singularities for PdSe$_2$, two for graphene, and one for MoS$_2$ and WS$_2$ monolayers. Most of these phase singularities are associated with $\Psi = 0°$ (equivalently $\rho = 0$ and, hence, $r_p = 0$), whereas one for graphene has $\Psi = 90°$ (equivalently $\rho = \infty$ or $r_s = 0$). Simulated angle-dependent spectrum of $\Delta$ in Figure 2 again demonstrates remarkable agreement with the experimental data, correctly predicting spectral positions of all phase singularities.

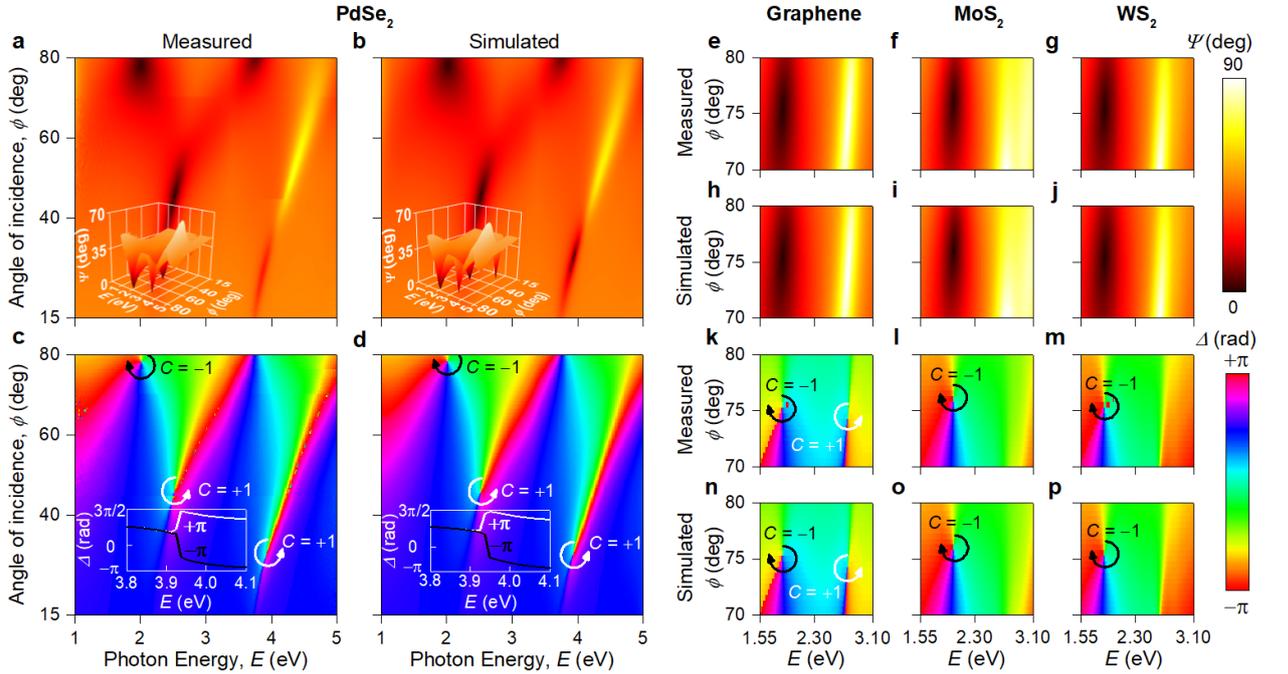

**Figure 2. Experimental observation of phase singularities. a, c** Experimental and **b, d** simulated ellipsometric parameters $\Psi$ (amplitude) and $\Delta$ (phase) for PdSe$_2$ (5.1 nm)/SiO$_2$(280 nm)/Si. The insets in panels (a) and (b) are their 3D view. The insets in panels (c) and (d) are phase behavior for incidence angles slightly above ($\phi$ = 30.5°) and below ($\phi$ = 30°) topological zero. **e-g, k-m** Experimental and **h-j, n-p** simulated $\Psi$ and $\Delta$ for graphene, MoS$_2$, and WS$_2$ on SiO$_2$/Si. In close vicinity of topological points, phase becomes singular and acquires topological charge $C = -1$ or +1. Optical constants of PdSe$_2$ for simulations are taken from Figure 3k. Meanwhile optical constants for graphene, MoS$_2$, and WS$_2$ were adopted from several reports.[34,35]

Previous works[4–6] realized these topological phase singularities only in metallic nanostructures through careful engineering of optical properties of nanostructured materials. Later, singular phase behavior was achieved in simple plasmonic heterostructures[36] where thin layers of metals (~20nm) and dielectric were used to generate zero reflection and phase singularities. Figure 2 proves that the heterostructure approach is quite general and could be realized even for 2D materials with ultimate atomic thickness in the simplest structure – 2D material/SiO$_2$/Si. The reason for this counterintuitive result is a rapid dielectric function variation (for example, due to excitons in TMDCs) in atomically thin HRIM. This rapid variation guarantees intersection with the zero-reflection surface (Figure 1b). Note that a thick layer is unsuitable for this purpose because absorption in that layer will prohibit interaction with the substrate's Fabry-Perot resonances.

To predict the position zero-reflection points for p- and s-polarized reflection in our structure shown in Figure 1a, we derived analytical expressions for the permittivities $\varepsilon_\text{p}$ and $\varepsilon_\text{s}$ of a thin film placed on a dielectric substrate which would result in the absence of the reflection for p- and s-polarized light, respectively (Supplementary Note 1):

$$\varepsilon_\text{p} = \frac{1}{k_0 t}\left(\frac{i\varepsilon_1}{q_{1z}} + \frac{\varepsilon_2}{q_{2z}}\frac{\frac{\varepsilon_3}{q_{3z}} - \frac{i\varepsilon_2}{q_{2z}}\tan(k_{2z}d)}{q_{3z}\tan(k_{2z}d) + \frac{i\varepsilon_2}{q_{2z}}}\right) \quad (2)$$

$$\varepsilon_\text{s} = \frac{1}{k_0 t}\left(iq_{1z} + q_{2z}\frac{q_{3z} - iq_{2z}\tan(k_{2z}d)}{q_{3z}\tan(k_{2z}d) + iq_{2z}}\right) \quad (3)$$

where $t$ and $d$ are thicknesses of the high-refractive-index material and dielectric layer respectively; $\varepsilon_1$, $\varepsilon_2$, and $\varepsilon_3$ are the dielectric permittivity of top halfspace, dielectric layer and bottom halfspace, in our case, that is air, SiO$_2$, and Si, respectively; $q_{iz} = k_{iz}/k_0 = \sqrt{\varepsilon_i - \varepsilon_1 \sin^2(\phi)}$ is the normalized z-component (perpendicular to layers) of the wavevector in medium number $i$, $k_0 = \omega/c$, $\omega$ is the frequency, $c$ is the speed of light, and $\phi$ is the angle of incidence. If the bottom halfspace is filled with a perfect electric conductor, the expressions can be simplified greatly to:

$$\varepsilon'_p = \frac{\varepsilon_2 \cot(k_{2z}d)}{q_{2z}k_0 t} \qquad \varepsilon''_p = \frac{\varepsilon_1}{k_{1z}t} \qquad (4)$$

$$\varepsilon'_s = \frac{q_{2z}\cot(k_{2z}d)}{k_0 t} \qquad \varepsilon''_s = \frac{q_1}{k_0 t} \qquad (5)$$

where $\varepsilon'_{p,s}$ and $\varepsilon''_{p,s}$ are the real and imaginary parts of dielectric permittivity $\varepsilon_{p,s}$.

Equations (2) and (3) define a zero-reflection surface in the parameter space of wavelength, real and imaginary parts of the permittivity ($\lambda$, Re[$\varepsilon$] and Im[$\varepsilon$], respectively). Intersections of the material dispersion curve in this parameter space with the zero-reflection surface of the system (thin film of HRIM/SiO$_2$/Si) define zero-reflection points for the particular material of the film, as shown in Figure 1b. The topology of mutual arrangement of the curve and the surface underlies the robustness of the zero-reflection effect to external perturbations (roughness, temperature change, etc.). If a perturbation is introduced to the thin film, a displacement of the material dispersion curve and/or the zero-reflection surface will only result in a shift of the zero-reflection point in the parameter space, but will not lead to its disappearance (Figure 1b). This argument is in line with non-trivial topological charges of the observed phase singularities: small changes in the parameters of the system cannot lead to a change in the phase round-trip around a point, since it is an integer of $2\pi$, thus making the zero-reflection point topologically protected.[37]

Spectral positions of topological phase singularities can be controlled by either the thickness or the dielectric permittivity of the material. The derived analytical expressions allow us to generalize the effect of phase singularities to other high-refractive-index materials (Supplementary Note 2). Hence, the effect of rapid phase change is universal for all atomically thin materials and substrates. It allows us hereafter to focus on PdSe$_2$, which demonstrates rich $\Psi$ and $\Delta$ spectra with a series of peaks and dips in Figure 2a-b. We begin with a detailed characterization of PdSe$_2$ film, and then switch to unique applications and features of topological phase gradient.

**Morphological and optical study of PdSe$_2$**

PdSe$_2$ thin films were prepared through chemical vapor deposition (CVD)[38] resulted in a uniform sample as confirmed by representative optical and scanning electron microscopy (SEM) images in Figure 3b-c. X-ray diffraction (XRD) spectrum showed pronounced peak in Figure 3b validating the high crystallinity of the film.[39] Next, we validated the material's purity by X-ray photoemission spectroscopy (XPS) in Figure 3g-h. It shows that Se:Pd atomic concentration ratio equals 1.92, close to the expected value of 2. Additionally, Raman spectra in Figure 3e-f have characteristic phonon modes $A_g^1$, $A_g^2$, $B_{1g}^2$, and $A_g^3$ inherent to PdSe$_2$ with puckered pentagonal crystal structure presented in Figure 3a.[40] This crystal configuration naturally has high geometrical and, therefore, high optical anisotropy (see Supplementary Note 3).[41,42]

To investigate anisotropic optical response, we measured Mueller matrices (Methods), which nonzero off-diagonal elements relate to sample anisotropy (see Supplementary Note 3). Interestingly, Mueller matrix' elements vary from point to point, as seen from Figure 3i-j. Conceivably, it comes from the random growth during the CVD synthesis since Mueller matrices' values in Figure 3i follow Gaussian law for random numbers. A similar random local anisotropic response is observed by polarized optical microscopy and Raman spectroscopy (Supplementary Note 3). Hence, at a macroscopic scale, our PdSe$_2$ layer exhibits an isotropic dielectric response. It allowed us to investigate optical constants by classical ellipsometric and reflectance measurements (see Supplementary Note 3) using the isotropic model for PdSe$_2$ with 5.1 nm thickness obtained by atomic force microscopy (AFM) in Figure 3d. The resulting broadband dielectric function is presented in Figure 3k. As expected, PdSe$_2$ has pronounced excitonic peaks[43] and a metallic Drude response caused by p-doping revealed by XPS (see Supplementary Note 3). Note that excitonic peaks of PdSe$_2$ align with Fabry-Perot resonances of the standard SiO$_2$ (280 nm)/Si substrate, making system PdSe$_2$/SiO$_2$(280 nm)/Si promising for enhancement of the PdSe$_2$ optical response.

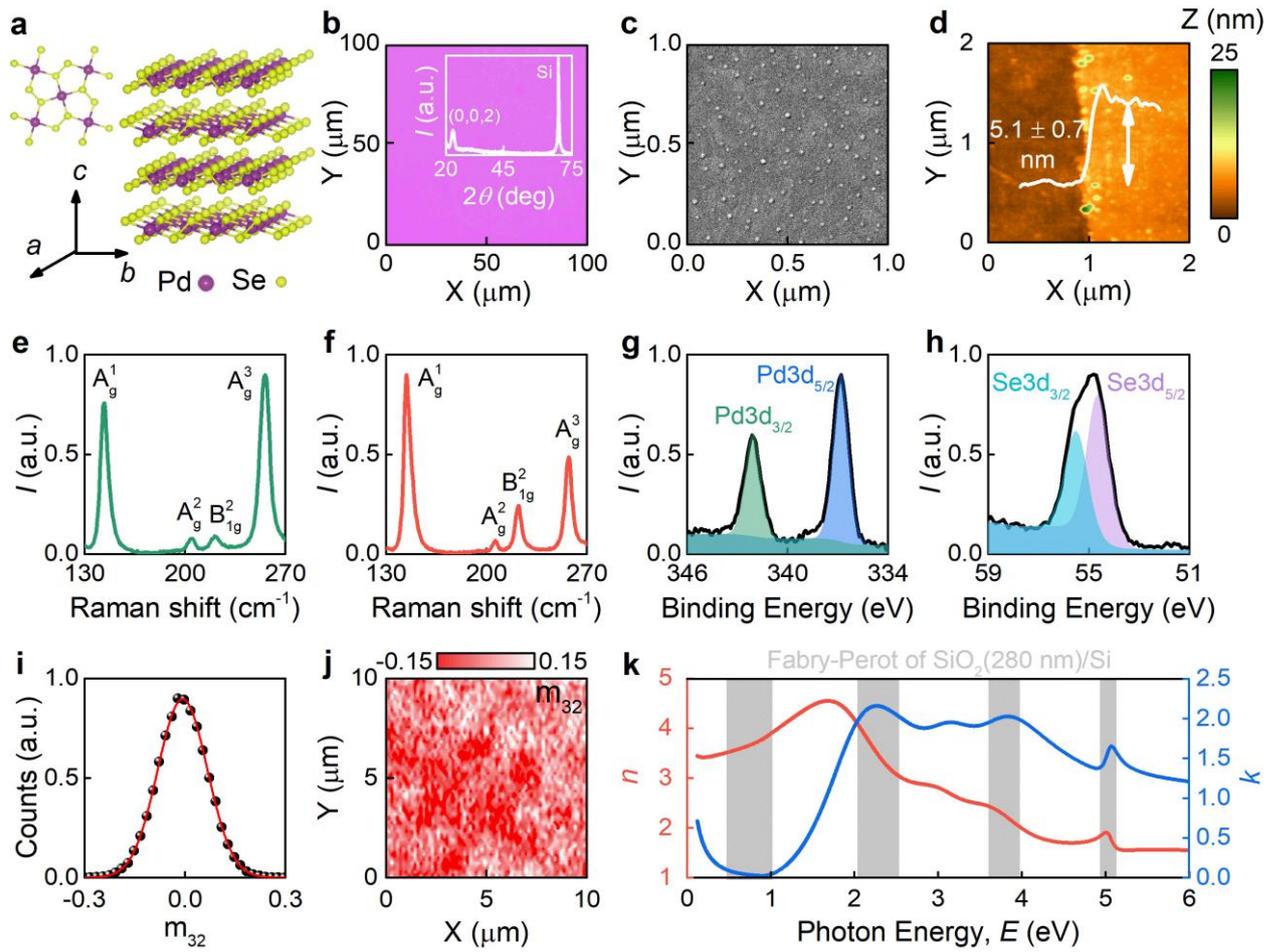

**Figure 3. Characterization of PdSe$_2$.** **a** Schematic illustration of PdSe$_2$ crystal structure. **b** Optical image of the sample. The inset shows PdSe$_2$ XRD diffraction pattern with pronounced peaks corresponding to the (0,0,2) crystal plane. Another peak is from Si substrate. **c** SEM image of the film. Small dots are seeding promoters for CVD growth. **d** AFM topography gives 5.1 nm film thickness. **e-f** Raman spectrum of PdSe$_2$ at excitation wavelengths $\lambda$ = 532 nm (green line) and 632.8 nm (red line). **g-h** XPS spectra of PdSe$_2$. **i-j** Relative frequency and map of $m_{32}$ (off-diagonal element of Mueller Matrix), indicating an anisotropic optical response. Although the material is highly anisotropic ($m_{32} \neq 0$), its random growth results in overall isotropic behavior (average $m_{32}$ = 0). The full Mueller Matrix is in Supplementary note 3. The red curve in panel **(i)** is a Gaussian fit. **k** Optical constants of PdSe$_2$ in the broad spectral range 0.124 – 6 eV (200 – 10000 nm). For PdSe$_2$ optical model, see Supplementary Note 4. Interestingly, the excitonic peaks of PdSe$_2$ coincide with the Fabry-Perot resonances of SiO$_2$ (280 nm)/Si.

**Applications of topological zeros: sensing**

Simple planar structures studied here are easy to incorporate and leverage in industrial and scientific devices where the optical phase plays a critical role.[5,24,25,31,44] The most prominent practical examples are holography,[24,25,44] image processing[45,46], label-free bio- or chemical sensing,[47–50] and quantum key distribution.[51,52] To validate the concept, we demonstrated that the liquid/$PdSe_2$/$SiO_2$/Si system is already an ultrahigh sensitive sensor owing to rapid phase change around the topological point. Note that for sensing measurements we used ellipsometer in the most accurate nulling mode (Methods) and 7.1 nm $PdSe_2$ thin film to have topological zero in the operation range of our device since liquid changes zero's spectral and angle position according to Equations (2-3).

For demonstration, we used water with 0, 2.5, 5, 15, and 20 % volume concentration of isopropanol. Notably, the measured $\Psi$ and $\Delta$ in water are in agreement with the predicted values (see Supplementary Note 5) whereby confirming the water stability of $PdSe_2$ in addition to its recently shown air stability.[40] Then, to alter the refractive index (RI), we injected isopropanol into the solution and recorded $\Psi$ and $\Delta$ (Figure 4a-b) for each water solution. As predicted, the change in amplitude response, $\Psi$, (Figure 4a and c) is relatively small due to the resonance's topological nature, whereas $\Delta$ (Figure 4b and d) shows a dramatic dependence on RI of liquid. Noteworthy, the phase sensitivity in our device of $7.5 \cdot 10^4$ degrees per refractive index unit (deg/RIU) exceeds that of the cutting-edge sensor based on plasmonic surface lattice resonance with $5.7 \cdot 10^4$ deg/RIU.[31] Therefore, the investigated system $PdSe_2$/$SiO_2$/Si is already a ready-to-use scalable device with outstanding performance thanks to the pronounced phase effect in topological points.

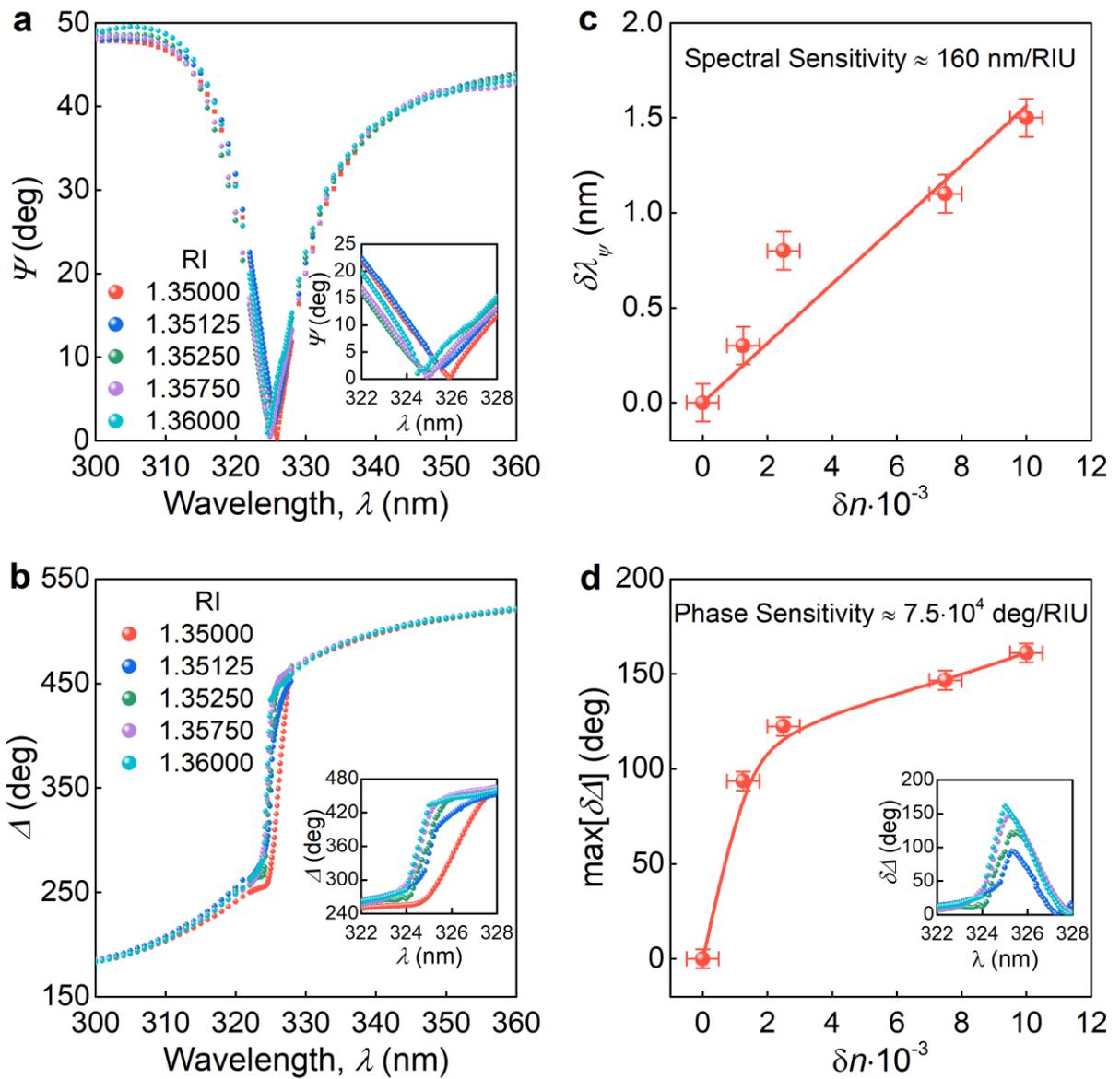

**Figure 4. Sensor based on topology of PdSe$_2$ film on SiO$_2$/Si. a, b** The dependence of ellipsometric parameters $\Psi$ (amplitude) and $\Delta$ (phase) on the refractive index (RI) of the investigated liquid recorded at the incidence angle $\theta = 49.4°$, corresponding to the topological zero. **c** Spectral shift of the resonance position of $\Psi$ spectrum with the change of the medium RI. **d** The maximum phase shift of the measured spectra with respect to the media with RI = 1.35 (water). The inset shows the phase shift of the measured spectra with respect to the media with RI = 1.35 (water).

**Evolution of phase singularities**

So far we have considered and observed reflection phase singularities with unitary topological charge, wherein the argument makes a $\pm 2\pi$ round-trip around the singularity. These points, however, are not stationary and evolve with the material parameters, as Equations (2-3) suggest. When two points with opposite charges (+1 and −1) meet in the parameter space, they annihilate leaving no phase singularity. We theoretically observe such annihilation with variation of the PdSe$_2$ film thickness, $t$, when two phase singularities with opposite topological charges meet at around $t \approx 3.2$ nm (Figure 5a). This case happens when the material dispersion curve in Figure 5b becomes tangent to zero-reflection surface. The corresponding angle-resolved spectrum of $\Delta$ shown in Figure 5b plotted for $t = 3.2$ nm reveals the absence of any phase singularities. Therefore, by varying the thickness one is able to control the position and amount of phase singularities, which appear or disappear in pairs, so that the total topological charge preserves.

Next, we examine the possibility of phase singularities with higher topological charges, $|C| > 1$. One potential opportunity for the emergence of a non-unitary topological charge is when a phase singularity either in $r_p$ or $r_s$ exhibits a non-unitary charge. Unfortunately, for non-magnetic materials in a planar system, zeros in $r_p$ or $r_s$ are essentially non-degenerate (Supplementary Note 6).

However, since $\rho$ is the ratio of two reflection coefficients, another possibility is when a $C = \pm 1$ phase singularity of $r_s$ coincides in the parameter space with a $C = \mp 1$ phase singularity of $r_p$. This gives rise to a $C = \pm 2$ phase singularity of $\rho$, which can be detected by spectroscopic ellipsometry. Equating the right-hand sides of Equations (2) and (3), we obtain an equation that determines, for a given substrate, the position of the point at which the zero-reflection conditions for both polarizations coincide; then we can immediately calculate the dielectric constant of the film that satisfies this condition. The charge of a point is determined by the derivative of the material dispersion curve, so the last step is to choose the direction of the material dispersion curve near the zero-reflection point. By using the set of parameters satisfying these conditions (Supplementary Note 6), we observe a $C = +2$ phase singularity in the spectrum of $\rho$, Figure 5c. The use of in-plane anisotropy can significantly assist in the design of non-unitary charged points of $\rho$. Suppose the main optical axis of the film is oriented along with the in-plane component of the wave vector, then the p-polarization is affected by only the longitudinal component of the permittivity and the s-polarization only by the transverse one. In that case, it is possible to select zero-reflection conditions for different polarizations entirely independently. Therefore, in contrast to isotropic materials, for which the position of points with a topological double charge depends on the properties of the substrate, in the case of in-plane anisotropic materials, it is potentially possible to obtain a point with $C = \pm 2$ at any point in the space of angles and frequencies (see Supplementary Note 7). One of such cases is shown in Figure 5d.

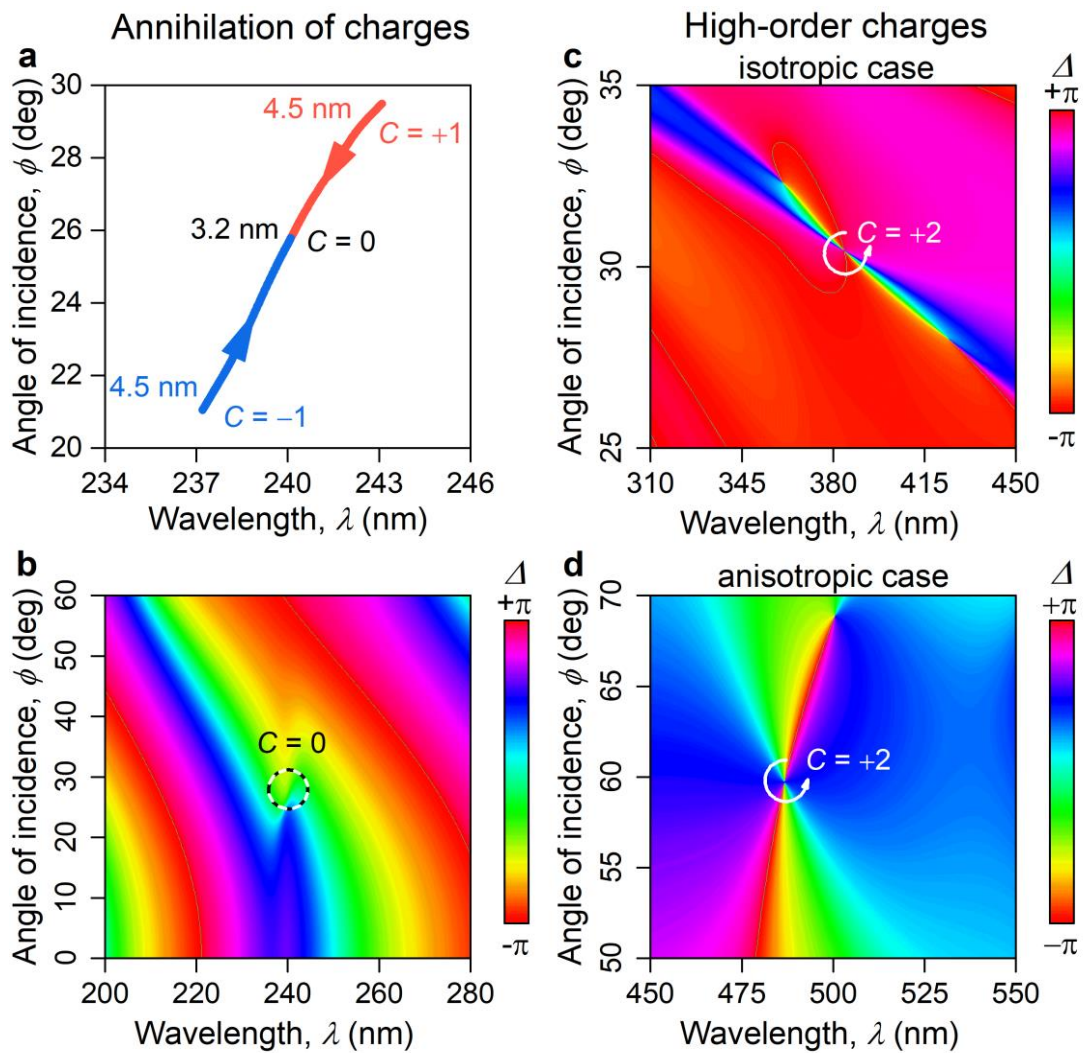

**Figure 5. Topological effects in phase singularities. a, b** The parameter change, here the thickness of PdSe$_2$, alters the positions of topological points, resulting, at specific thickness ($t = 3.2$ nm), in annihilation of opposite topological charges. **c, d** Topological charges with $C = +2$ for ellipsometric phase $\varDelta$ with isotropic and anisotropic thin films with system's parameters collected in Supplementary Note 7.

For $C = \pm 2$ singularities demonstrated in Figure 5c-d, the phase makes a $\pm 4\pi$ round-trip around such a singularity, thus increasing the local phase gradient approximately by a factor of two, which could significantly improve phase applications. For instance, the sensitivity of the corresponding refractive index sensor may increase approximately two-fold.

The scenario of non-unitary phase singularities is somewhat reminiscent of exceptional points in non-Hermitian optical systems.[53] Such points, which correspond to coalescent eigenstates of a non-Hermitian Hamiltonian, feature a strong $(\epsilon - \epsilon_0)^{1/(N+1)}$ dependence of the eigenenergies of the optical system on a perturbation parameter $\epsilon$ in the vicinity of the exceptional point $\epsilon_0$ (with $N$ being the order of the exceptional point), which has been used for boost the sensitivity.

Additionally, upon appropriate engineering of the system the same concept of topological phase manipulation could also be applied in transmission regime, thus bridging our phase engineering approach with metasurfaces.

**DISCUSSION**

Flat optics enable the design of optical components into thin, planar, and CMOS-compatible structures. Coupled with 2D materials, it evolves into 2D flat optics with ultracompact and tunable devices. Nevertheless, atomically thin optical elements suffer from low efficiency of phase manipulation. To lift this limitation and achieve phase control with 2D materials, we utilized topologically protected zeros of a simple heterostructure. We showed both experimentally and theoretically that a whole set of high-index 2D materials could provide rapid phase variations revealed by spectroscopic ellipsometry. In addition, we demonstrate that topological approach leads to high-performance devices on the sensing example and propose the future direction of topological phase effects such as annihilation and high-order charges. From a broader perspective, our results open new avenues for effective application of atomically thin high-refractive-index materials as phase materials in photonics.

**ACKNOWLEDGMENTS**

G.E., K.V., D.G.B., G.T., D.Y., S.N., A.V., A.M., I.K., A.A., and V.V. gratefully acknowledges the financial support from the Ministry of Science and Higher Education of the Russian Federation (Agreement No. 075-15-2021-606). A.N.G. acknowledge EU Graphene Flagship, Core 3 (881603).

**AUTHOR CONTRIBUTIONS**

†These authors contributed equally. A.N.G., V.V., K.S.N., V.K., and A.A. suggested and directed the project. G.E., V.K., G.T., Y.S., D.Y., S.N., S.Z., R.R., and A.M.M. performed the measurements and analyzed the data. K.V., D.G.B., G.E., A.M., I.K., and A.V. provided theoretical support. G.A.E., K.V., and D.G.B. wrote the original manuscript. G.E., K.V., D.G.B., A.N.G., K.S.N., V.V., and A.A. reviewed and edited the paper. All authors contributed to the discussions and commented on the paper.

**COMPETING INTERESTS**

The authors declare no competing interests.

**METHODS**

**Ellipsometry measurements.** For visualization of topological charge in phase, we used a variable-angle spectroscopic ellipsometer (VASE, J.A. Woollam Co.) since it measures amplitude and phase of complex reflection ratio simultaneously. Measurements were done over a wide wavelength range from 248 to 1240 nm (1 – 5 eV) in steps of 1 nm and multiple angles of incidence in the range 15-80° with 0.5° step. Ellipsometer has a single chamber monochromator with two gratings: 1200 g/mm for visible light (248 – 1040 nm) with 4.6 nm bandwidth and 600 g/mm for near-infrared interval (1040 – 1240 nm) with 9.2 nm bandwidth.

For accurate refractive index sensing, we performed the nulling ellipsometry with Accurion nanofilm_ep4 ellipsometer. During the measurement light was initially directed through the polarizer then through the compensator, whose settings were adjusted until the reflection from the sample became linearly polarized. Afterward, the analyzer was set to achieve the minimum in the signal at the photodetector. The final positions of polarizer and analyzer at signal's minimum uniquely define ellipsometric parameters $\Psi$ and $\Delta$. Measurements were done over a wavelength range from 300 to 360 nm in steps of 0.1 nm with 4 nm bandwidth and at 49.4° incident angle corresponding to the singular point of $PdSe_2$ in water.

To probe anisotropic response, we also measured 11 elements of Mueller Matrix ($m_{12}$, $m_{13}$, $m_{21}$, $m_{22}$, $m_{23}$, $m_{24}$, $m_{31}$, $m_{32}$, $m_{33}$, $m_{34}$) on Accurion nanofilm_ep4 ellipsometer over 400 – 1000 nm wavelength range in 5 nm step with 4 nm bandwidth at 50° incident angle in a rotation compensator mode to get access to Stokes parameters in the input branch of ellipsometer.

**Optical visualization.** The surface images (2400 × 2400 pixels) of $PdSe_2$ were captured by an optical microscope (Nikon LV150L) with a digital camera DS-Fi3.

**Atomic force microscopy.** The thickness and surface morphology of the $PdSe_2$ film were accurately characterized by an atomic force microscope (AFM, NT-MDT Spectrum Instruments "Ntegra") using AFM in a peak-force mode at ambient conditions. AFM measurements were carried out using cantilever tips from NT-MDT (ETALON, HA_NC) with a spring constant of 3.5 N/m, a tip radius < 10 nm and a resonant frequency of 140 kHz. Images of $PdSe_2$ surface were taken with 512 × 512 pixels and scan rate of 0.5 Hz, after that data were analyzed by Gwyddion software.

**Scanning electron microscopy.** Scanning electron microscopy (SEM, JEOL JSM-7001F) with a Schottky emitter in secondary electron imaging mode with a voltage of 30 kV and current of 67 µA, and a working distance of ~ 6.3 mm was applied to study in detail surface features and homogeneity of the PdSe$_2$ film surface within different areas using 1960 × 1280 pixel scan.

**XPS characterization.** The chemical states of the elements in the film, as well as the valence band were analyzed by X-ray photoelectron spectroscopy (XPS) in Theta Probe tool (Thermo Scientific) under ultrahigh vacuum conditions (base pressure < $10^{-9}$ mBar) with a monochromatic Al-K$_α$ X-ray source (1486.6 eV). Photoelectron spectra were acquired using fixed analyzer transmission (FAT) mode with 50 eV pass energy. The spectrometer energy scale was calibrated on the Au 4f$_{7/2}$ line (84.0 eV).

**X-ray diffraction.** X-ray powder diffractometrer (XRD, Thermo ARL X'TRA) equipped with Cu K$α$ radiation $λ = 0.154$ nm was used to characterize the crystalline structure and phase of PdSe$_2$ film. The XRD pattern was taken at ambient conditions by 2$θ$-scan over the range of 20–75° with a step of 0.05° and accumulation time of 2 s.

**Reflectance measurements.** Fourier-transform spectrometer Bruker Vertex 80v has been used to measure the reflection coefficient at the normal incident at room temperature in the frequency range from 1000 to 24000 cm$^{-1}$, as the reference was used the reflection from the gold mirror.

**Raman characterization.** The experimental setup used for Raman measurements was a confocal scanning Raman microscope Horiba LabRAM HR Evolution (HORIBA Ltd., Kyoto, Japan). All measurements were carried out using linearly polarized excitation at wavelengths 532 and 632.8 nm,1800 lines/mm diffraction grating, and ×100 objective (N.A. = 0.90), whereas we used un-polarized detection to have a significant signal-to-noise ratio. The spot size was approximately 0.43µm. The Raman images were recorded by mapping the spatial dependence of Raman intensity integrated at the main Raman peaks within the shift range 136 – 156 cm$^{−1}$, for each of the 45 × 45 points in the scan, with an integration time of 500 ms at each point and incident power P = 1.7 mW.

**DATA AVAILABILITY**

The datasets generated during and/or analyzed during the current study are available from the corresponding author on reasonable request.